\documentclass[12pt,preprint]{aastex}

\shortauthors{Wright}

\shorttitle{Cosmology Calculator}

\newcommand{\etal}         {{\it et al.}}

\newcommand{\be}           {\begin{equation}}
\newcommand{\ee}           {\end{equation}}
\newcommand{\bea}           {\begin{eqnarray}}
\newcommand{\eea}           {\end{eqnarray}}

\slugcomment{Draft: 6 Oct 2006}

\begin{document}

\title{A Cosmology Calculator for the World Wide Web}

\author{
E.\ L.\ Wright\altaffilmark{1},
}

\altaffiltext{1}{UCLA Astronomy, PO Box 951547, Los Angeles CA 90095-1547, USA}

\email{wright@astro.ucla.edu}

\begin{abstract}
A cosmology calculator that computes times and distances as a function of
redshift for user-defined cosmological parameters is available at
http://www.astro.ucla.edu/$\sim$wright/CosmoCalc.html.
This note gives the formulae used by the cosmology calculator and discusses
some of its implementation.
A version of the calculator that allows one to specify the equation of
state parameter $w$ and $w^\prime$ is at ACC.html, and a version for
converting the light travel times usually given in the popular press into
redshifts is at DlttCalc.html.
\end{abstract}

\keywords{cosmology: miscellaneous -- methods: miscellaneous}

\section{INTRODUCTION}\label{intro}

There are many definitions of distances in cosmology, and it is often
frustrating to calculate the kind of distance needed for a given
situation.  In particular, press releases (Clavin, 2006) 
very often use the {\em light
travel time distance}, $D_{ltt} = c(t_\circ-t_{em})$, even though this
distance has very many undesirable characteristics.  I wrote the
cosmology calculator in order to be able to calculate $D_{ltt}$ along
with the observable angular and luminosity distances $D_A$ and $D_L$.
It also computes the proper radial distance which is the only distance
that is compatible with the common definition of distance for moving
objects:  the spatial separation at a common time.  The common time
used is now, or $t_\circ$.  This it is called $D_{now}$ below.
$D_{now}$, also known as the comoving radial distance, is not measurable from
a single position, but it can in principle be measured by a family of
comoving observers.  It is the one distance that satisfies the Hubble
Law exactly: $dD/dt = H(t)D$ without any corrections.  Other distances
are defined using quantities taken from more than one time and
generally only satisfy the Hubble Law to first order.

\section{EQUATIONS \label{sec:obs}}

The following results have been given many times before.
Hogg (1999) gives equations equivalent to the ones in this paper.
But for completeness in documenting the cosmology calculator
the equations are spelled out below.

The metric is given by
\be
ds^2 = c^2dt^2 - a(t)^2 R_\circ^2 (d\psi^2 + S^2(\psi)[d\theta^2+
	\sin^2\theta d\phi^2])
\ee
where $S(x)$ is $\sinh(x)$ if $\Omega_{tot} < 1$ and $\sin(x)$ for
$\Omega_{tot} > 1$.  The radius of curvature is given by
$R_\circ = (c/H_\circ)/\sqrt{|1-\Omega_{tot}|}$
since I use the normalization that $a = 1$ at the current time so
the redshift is given by $1+z = 1/a$.
The past light cone is given by $c dt = a(t) R_\circ d\psi$ so
the comoving radial distance is
\be
D_{now}  =  R_\circ \psi = \int \frac{c dt}{a}
= \int_{1/(1+z)}^1 \frac{c da}{a\dot{a}}
\ee
and of course the light travel time distance is given by
\be
D_{ltt} = \int c dt = \int_{1/(1+z)}^1 \frac{c da}{\dot{a}}
= c(t_\circ - t(z))
\ee
The exact dynamics of the Universe can be captured using the
the energy equation from the Newtonian approximation to
cosmology, which gives
\be
\frac{\dot{a}^2}{2} - \frac{G M(<a)}{a} = \mbox{const}.
\ee
GR modifies the acceleration equation by including the pressure
as a source of gravity, but this just cancels the variation of
the enclosed mass $M(<a)$ caused by the pressure, so the energy
equation is exact.
So we can write $\dot{a}$ as $H_\circ \sqrt{X}$ with
\be
X(a) = \Omega_m/a + \Omega_r/a^2 + \Omega_v a^2 + (1-\Omega_{tot})
\approx 1 + 2q_\circ z + \ldots
\ee
where $q_\circ$ is the deceleration parameter.
Let us define a quantity $Z$ that is the comoving distance divided by
the Hubble radius $c/H_\circ$.  Then
\be
Z = \int_{1/(1+z)}^1 \frac{da}{a\sqrt{X}} = \int_0^z \frac{dz}{(1+z)\sqrt{X}}
\approx z - z^2(1+q_\circ)/2 + \ldots
\ee
To first order this agrees with the redshift, so $Z = z+ \ldots$
This definition gives
$D_{now} = (cZ/H_\circ)$ and
$D_{ltt} = (c/H_\circ) \int_{1/(1+z)}^1 da/\sqrt{X}$.
Then the angular size distance, which is defined as $D_A = R/\theta$
where $R$ is the transverse proper length of an object that subtends
an angle $\theta$ on the sky, is given by:
\bea
D_A & = & \frac{c}{H_\circ}
\frac{S(\sqrt{|1-\Omega_{tot}|}Z)}{(1+z)\sqrt{|1-\Omega_{tot}|}}
\nonumber \\
& = & \frac{D_{now}}{(1+z)}
\left(1 + \frac{1}{6} (1-\Omega_{tot}) Z^2 + \frac{1}{120} 
(1-\Omega_{tot})^2 Z^4 + \ldots \right)
\eea
We can define 
a function $J(x)$ given by
\be
J(x) =  \cases{
\sin(\sqrt{-x})/\sqrt{-x}, & $x < 0$;\cr
\sinh(\sqrt{x})/\sqrt{x}, & $x > 0$;\cr
1 + x/6 + x^2/120 + \ldots + x^n/(2n+1)! + \ldots, & $x \approx 0$.\cr
}
\ee
Then
\bea
D_{ltt} & = & \frac{c}{H_\circ} \int_{1/(1+z)}^1 \frac{da}{\sqrt{X}}
\nonumber \\
D_{now} & = & \frac{c}{H_\circ} \int_{1/(1+z)}^1 \frac{da}{a\sqrt{X}} = 
\frac{cZ}{H_\circ}
\nonumber \\
D_A & = & D_{now} \frac{J([1-\Omega_{tot}]Z^2)}{1+z}
\nonumber \\
D_L & = & (1+z)^2 D_A
\eea
Note that the luminosity distance $D_L$ is defined to make the inverse
square law work perfectly for bolometric fluxes, so that 
$F_{bol} = L/(4\pi D_L^2)$ for an object of luminosity $L$.

The enclosed volume calculation requires the integral of
either $\sin^2\psi d\psi$ or $\sinh^2\psi d\psi$.  The
closed case becomes
\be
V(<z) = 4\pi R_\circ^3 \int_0^\psi \frac{1-\cos 2\psi^\prime}{2} d\psi^\prime
= R_\circ^3 \left(\frac{\psi}{2}- \frac{\sin 2\psi}{4} \right)
\ee
Note that $\psi = \sqrt{|1-\Omega_{tot}|}Z(z)$.  The open case gives
\be
V(<z) = 4\pi R_\circ^3 \left(\frac{\sinh 2\psi}{4} - \frac{\psi}{2}\right)
\ee
The ratio of $V(<z)$ to the Euclidean formula $(4\pi/3)(R_\circ \psi)^3$ is
given by
\bea
\frac{V(<z)}{(4\pi/3)(R_\circ \psi)^3} & = &
1+\sum_{n=1}\frac{6\times4^n}{(2n+3)!}[(1-\Omega_{tot})Z(z)^2]^n
\nonumber \\
& = & 1 +\frac{1}{5}[(1-\Omega_{tot})Z(z)^2]
+\frac{2}{105}[(1-\Omega_{tot})Z(z)^2]^2 + \ldots
\eea
in both the open and closed cases.

\section{IMPLEMENTATION\label{sec:analysis}}

The cosmology calculator is implemented as a Web page 
(CosmoCalc.html) that has
a large number of Javascript definitions in the header,
followed by immediately executed Javascript that writes a
frameset to the current document.  The frameset calls 
for the input form (CCform.html) and the output page
(CCout.html).  If Javascript is not enabled, or if there
is an error in the Javascript, then the body of CosmoCalc.html
is displayed.  This body is just an error message saying that
Javascript must be enabled.

I have received several requests for the code used in the cosmology
calculator.  But since the code is in Javascript, it is in the HTML
files in ASCII form.  
It is easy to save the page source to get the code, and it
is easy to modify the code using any text editor.  If your modifications
introduce an error, then you will see the error message saying that
Javascript must be enabled.  This only means that you must find the error
in your modified CosmoCalc.html.  I have had to do this dozens of times,
so don't be discouraged.

Even if you do not intend to modify the code, downloading the three
HTML files will let you run the calculator locally when not connected
to the Internet.

The numerical evaluation of the integrals for $Z$ and $t$ is done using
the mid-point rule with 1000 panels.  While this is not a very efficient
use of CPU time, it was very easy to code.  And with Javascript, the CPU
time is consumed on the user's computer instead of the server.
The functions being integrated go smoothly to zero as $a$ goes to zero
as long as $\Omega_r > 0$.

Another hidden aspect of the cosmology calculator is that it
automatically sets the radiation density $\Omega_r h^2$ to the
value appropriate for $T_\circ = 2.72528$\,K and three massless
neutrino species, $\Omega_r h^2 = 4.165 \times10^{-5}$.  Here $h
= H_\circ/100\;\mbox{km/sec/Mpc}$, and this factor includes a small
($< 1\%$) boost in the neutrino density due to a slight transfer to
$e^+e^-$ annihilation energy into neutrinos (Hannestad \& Madsen, 1995).  
So if you want to verify some
simple cases, like the empty Universe, you should use a large value of
the Hubble constant which reduces the relative importance of this
radiation term.  For example, the open button with $H_\circ = 97.78$
and $\Omega_M = 0$ gives an age of the Universe of 9.934 Gyr which is
0.7\% from the expected 10 Gyr.
But using $H_\circ = 977.8$ gives an age of 999 Myr which
is only 0.1\% from the expected 1\,Gyr.

For very early times, the age of the Universe is adjusted to allow
for the existence of $e^+,e^-$, $\mu^+,\mu^-$, etc. using the
$g_*$ and $g_{*S}$ for the standard model of particle physics given
by Kolb \& Turner (1990).

\section{Light travel time inversion}

A slightly modified version of the cosmology calculator is at\\
http://www.astro.ucla.edu/$~\sim$wright/DlttCalc.html.\\
The input form asks for the light travel time in Gyr instead of redshift.
The redshift is found by evaluating the integral for light travel time
in steps of $-0.001$ in $a$ starting from $a=1$
until the input value is exceeded, and then
interpolating to get $a$ and thus $z$.  Once $z$ is known the calculations
proceed as in CosmoCalc.html.

\section{MORE OPTIONS\label{sec:discussion}}

Since the cosmology calculator was first written, there have been two
developments that change the kinds of models that people want to run.
I have created a new version of the cosmology calculator with more
options and placed it in\\
http://www.astro.ucla.edu/$\sim$wright/ACC.html\\
One development is the discovery of neutrino oscillations, indicating
that the assumption of massless neutrinos is not correct.  The neutrino
temperature is $(4/11)^{1/3} T_\circ = 1.95$\,K, and a typical momentum
for a thermal neutrino is $\approx 3 kT/c$.  This corresponds to the rest
energy of a neutrino with $mc^2 \approx 0.0005$\,eV.  Since the neutrinos
thermalized while still relativistic, their distribution is that of a
relativistic Fermi-Dirac particle, so the neutrino density is given by
\bea
\rho_\nu & = & 4\pi g_s h^{-3}
\sum_{e,\mu,\tau}
\left(\int \sqrt{m_\nu^2 + p^2/c^2} \; \frac{p^2 dp}{\exp(pc/kT_\nu)+1}\right)
\nonumber \\
& = & 4\pi g_s \left(\frac{kT_\nu}{hc}\right)^3 \sum_{e,\mu,\tau}
\left(\int \sqrt{m_\nu^2 +(x kT_\nu)^2/c^4}\;\frac{x^2 dx}{\exp(x)+1}\right)
\eea
The number of spin states for a neutrino is one, but to allow for 
anti-neutrinos one should use $g_s = 2$.  

The integral over neutrino momentum has to be evaluated for each step of
the integration over $a$ so it needs to be done efficiently, even when the
work is done on the user's computer.  In the low temperature limit when
$kT_\nu << m_\nu c^2$ the integral over $x$ evaluates to $m_\nu
(3/4)\zeta(3)\Gamma(3)$, while in the high temperature limit $kT_\nu >> m_\nu
c^2$ the integral evaluates to $(k T_\nu/c^2)(7/8)\zeta(4)\Gamma(4)$.
Both limits can be evaluated correctly by approximating the integral
using 
\be 
\int \sqrt{m_\nu^2 + x^2(kT_\nu)^2/c^4}\;\frac{x^2
dx}{\exp(x)+1} \approx m_\nu (3/4)\zeta(3)\Gamma(3) \sqrt{1+(x_1
kT_\nu/m_\nu c^2)^2} 
\ee 
with the single integration knot at $x_1 =
[(7/8)\zeta(4)\Gamma(4)]/[(3/4)\zeta(3)\Gamma(3)] = 3.151$.  This
approximation has a maximum error of $< 3\%$.  But a better
approximation is 
\be 
\int \sqrt{m_\nu^2 + x^2(kT_\nu)^2/c^4}\;\frac{x^2 dx}{\exp(x)+1} 
= m_\nu (3/4)\zeta(3)\Gamma(3) f(x_1 kT_\nu/m_\nu c^2)
\ee 
with 
\be 
f(y) \approx (1+y^\beta)^{1/\beta} 
\ee 
and $\beta = 1.842$ which
has a maximum error $< 0.3\%$.  The mass at which neutrinos are
semi-relativistic at the current epoch is 
\be 
m_{rel}c^2 = x_1 kT_\nu = 0.000531(T_\circ/2.72528\;\mbox{K})\;\mbox{eV}
\label{eq:mrel}
\ee 
The final result is that the effective neutrino
density can be written as 
\be 
\Omega_\nu(z) h^2 = \left(\frac{T_\circ}{2.72528\;\mbox{K}}\right)^3
\frac{\sum_{e,\mu,\tau} m_\nu f(m_{rel}(1+z)/m_\nu)}
{93.14\;\mbox{eV}/c^2} 
\ee
The normalization of 93.14~eV is from Mangano \etal\ (2005) and is 1.05\%
higher than the nominal due to residual coupling of annihilation energy
to the neutrinos.  In the relativistic limit the density is 1.53\% higher.
Increasing $T_\nu$ by 0.48\% over the nominal $(4/11)^{1/3}T_\circ$ when
computing $m_{rel}$ allows for the relativistic limit, and this boost is
included in Eq(\ref{eq:mrel}).

This $\Omega_\nu(z)$ is just a function of $z$ which gives $\Omega_\nu$
at $z=0$ and should not be confused with the actual $\Omega_\nu$ at $z$'s
other than zero.
In ACC.html the neutrino density at $z=0$ is
subtracted from the input $\Omega_M$ giving separate $\Omega_{CM} =
\Omega_M - \Omega_\nu(z=0)$ for the CDM and baryons,
and $\Omega_\nu$ for the neutrinos.
Of course the neutrinos are not included in the radiation so
$\Omega_r h^2 = 2.778 \times 10^{-5} (T_\circ/2.72528)^4$.

For a hierarchical neutrino mass pattern, with masses $\approx 0.001,
\;0.009\;\&\; 0.049$~eV, the change in distances introduced by
neutrino masses is negligible.  At redshifts up to 5 in the
WMAP concordance flat $\Lambda$CDM
cosmology with $\Omega_M = 0.27$ and $H_\circ = 71$, the
changes are less than 0.01\%.   This mass pattern is the default
when ACC.html is loaded.  But even the more massive nearly
degenerate neutrino mass patterns such as $\approx 0.33,\;
0.33\;\&\; 0.33$~eV have a minimal effect on the
distances and times.

\begin{figure}[t]
\plotone{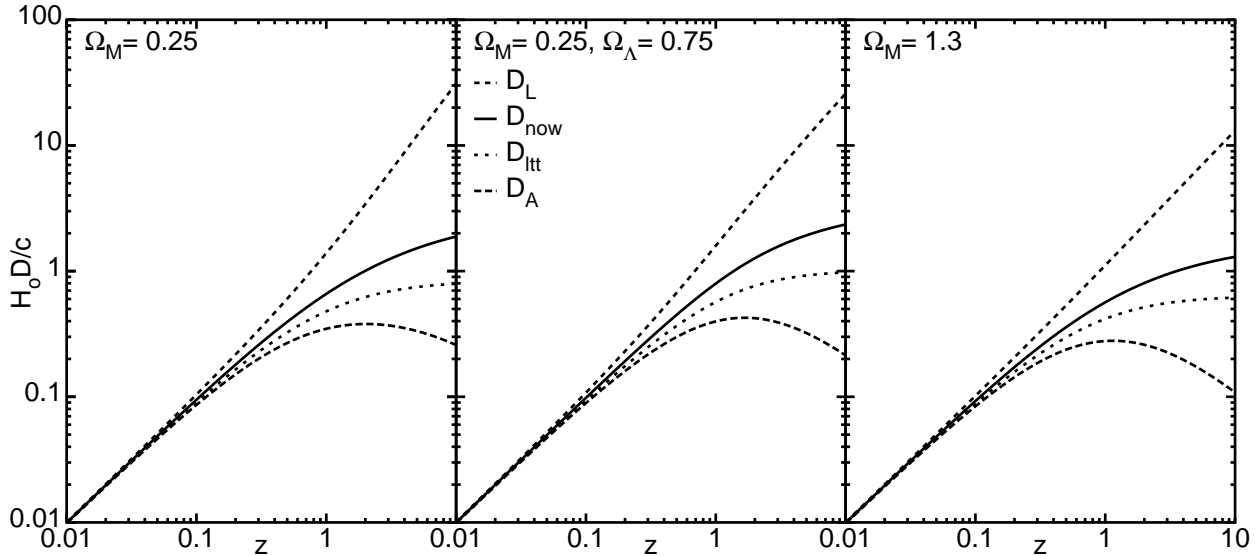}
\caption{The 4 distance measures $D_A$, $D_{ltt}$, $D_{now}$
and $D_L$ from bottom to top plotted against a fifth distance measure:
the redshift $z$, for 3 different cosmological
models:  an open CDM model, a flat $\Lambda$CDM model, and a closed CDM model.
The center and right panels are consistent with the CMB angular power spectrum,
while the left and center panels are consistent with large scale structure.
Only the center panel is consistent with supernova data.\label{fig:Dsvsz}}
\end{figure}

The second change in the paradigm is the introduction of the equation
of state $w(z) = P/\rho c^2$ as a parameter in the model.  I have
implemented $w(z)$ as $w = w_\circ+2w^\prime(1-a)$ following Linder
(2003) who added the factor of ``2'' normalization
to the simple form used by Chevallier \& Polarski (2001).
This functional form behaves well at high redshift and it
allows an analytic solution for the dark energy density as a function
of $z$.  This solution is
\be
\rho_{DE} = \rho_{DE,\circ} \; a^{-(3+3w_\circ+6w^\prime)} \; 
\exp(-6w^\prime[1-a])
\ee
The defaults are $w = -1$ and $w^\prime = 0$ and in that case
ACC.html reduces to the CosmoCalc.html case.

Unlike the neutrino masses, changes in $w$ have substantial effects on
distances and ages.  Changing $w$ to $-0.7$ instead of the default $-1$
changes the age of the Universe by 6\% and the luminosity distance at $z=1$
by 7\% when $\Omega_M = 0.27$, $\Omega_{DE} = 0.73$, and $H_\circ = 71$
are left unchanged.
But for $w=-0.7$ the model consistent with both supernovae (Riess \etal\
2004 gold+silver) and the CMB (Page \etal\ 2003 peak positions)
changes to an open model with $\Omega_M = 0.19$, $\Omega_{DE} = 0.74$ and
$H_\circ = 84$, and this gives an age change of 17\% and a luminosity distance
change of 21\% at $z=1$.  However, if $\Omega_M$ and $\Omega_{DE}$ are allowed
to vary as free parameters, the observable supernova signal in $D_L$ is reduced 
to only 0.3\% (Wright 2005). 

Finally ACC.html allows one to input $T_\circ$, allowing for easier tests
of simple cases.

With these changes, the $X$ function is
\be
X(a) = (\Omega_{CM}+\Omega_\nu(z))/a + \Omega_{r}/a^2
        + \Omega_{DE}\;a^{-(1+3w_\circ+6w^\prime)}\;\exp(-6w^\prime[1-a])
        + (1-\Omega_{tot})
\ee
The rest of the calculation of distances is unchanged.

\section{CONCLUSION\label{sec:conclusion}}

These cosmology calculators are suitable for interactive use, providing fairly
quick answers for single cases.  Users who wish to use the code for large scale 
calculations should translate it into a compiled language and change the
quadrature formula.  In particular, if fitting to datasets with redshifts and
distance, the data should be sorted by redshift and the distance integrals
evaluated for all objects in a single pass through the sorted data.  Plotting 
figures is an obvious case where redshifts are computed in order: 
Figure \ref{fig:Dsvsz} shows the distance
measures discussed in this paper as a function of $z$ for 3 different 
models.  The formulae presented here were translated into Postscript 
for this figure, resulting in a 4 kB file.

\acknowledgements

\thebibliography

\bibitem{chevallier/polarksi;2001} Chevallier, M. \& Polarski, D. 2001,
Int. J. Mod. Phys., D10, 213.

\bibitem{clavin;2006}
Clavin, W. 2006, JPL press release,\\ 
http://www.nasa.gov/centers/jpl/news/spitzerf-20060321.html

\bibitem{hannestad/madsen;1995}
Hannestad, S. \& Madsen, J. 1995, PRD, 52, 1764-1769.

\bibitem{hogg;1999} Hogg, D. 1999, astro-ph/9905116v4

\bibitem{kolb/turner;1990} Kolb, E. \& Turner, M. 1990,
``The Early Universe'', (Addison Wesley)

\bibitem{linder;2003} Linder, E. 2003, PRL, 90, 091301

\bibitem{mangano/etal;2005} Mangano, G., Miele, G., Pastor, S., Pinto, T., 
Pisanti, O. \& Serpico, P. 2005, Nucl.Phys. B, 729, 221-234. 

\bibitem{page/etal;2003}
Page, L. \etal, 2003, ApJS, 148, 233. 

\bibitem{riess/etal;2004}
Riess, A. \etal., 2004, \apj,  607, 665-687. 

\bibitem{wright;2005}
Wright, E. 2005, New Astronomy Reviews, 49, 407.

\end{document}